# SmartCoAuth: Smart-Contract privacy preservation mechanism on querying sensitive records in the cloud

Muhammed Siraj[1*], Mohd. Izuan Hafez Hj. Ninggal[1], Nur Izura Udzir[1], Muhammad Daniel Hafiz Abdullah[1], Aziah Asmawi[1]

[1]*Faculty of Computer Science & Information Technology, UPM, Serdang, Selangor, Malaysia*

*\*salisiraj@gmail.com*

**ABSTRACT**

Sensitive records stored in the cloud such as healthcare records, private conversation and credit card information are targets of hackers and privacy abuse. Current information and record management systems have difficulties achieving privacy protection of such sensitive records in a secure, transparent, decentralized and trustless environment. The Blockchain technology is a nascent and a promising technology that facilitates data sharing and access in a secure, decentralized and trustless environment. The technology enables the use of smart contracts that can be leveraged to complement existing traditional systems to achieve security objectives that were never possible before. In this paper, we propose a framework based on Blockchain technology to enable privacy-preservation in a secured, decentralized, transparent and trustless environment. We name our framework SmartCoAuth. It is based on Ethereum Smart Contract functions as the secure, decentralized, transparent authentication and authorization mechanism in the framework. It also enables tamper-proof auditing of access to the protected records. We analysed how SmartCoAuth could be integrated into a cloud application to provide reliable privacy-preservation among stakeholders of healthcare records stored in the cloud. The proposed framework provides a satisfactory level of data utility and privacy preservation.

I. INTRODUCTION

Due to the huge number of data breaches experienced in organizations such as the healthcare industry, researchers have since proposed and implemented several techniques in order to mitigate the attacks. Protecting the privacy of sensitive records of patients which are kept by a provider in the cloud is of high priority due to the negative effect it can cause to the victims and other stakeholders. In the case of healthcare, these records could be Personal Health Records (PHR) of patients or general Electronic Healthcare Records (HER).

Privacy in computer security generally refers to the control that data subjects or owners have on their records and how they are able to influence what information is collected and stored about them as well as by whom and to whom the said information is disclosed [1]. With respect to Healthcare records and the requirement of HIPAA, the major goal is to properly protect the health information while at the same time allowing the sharing of the records needed for providing quality healthcare services as well as protect the wellbeing of the public [2].

The abuse on EHR also has financial implications due to the fines that come as a result of violating the privacy requirement. Maintaining the privacy of HER is therefore paramount and of high priority, to the provider and other stakeholders because the abuse of the privacy requirements of the records can have detrimental effects and consequences. Their goal is to prevent access to such private records by unauthorized users.

Due to over-dependence on traditional system architecture, attackers have various and sometimes easier ways to subvert the system in other to breach the data privacy or security of sensitive records such as EHR stored in the cloud. Given such traditional architecture, the security and protection of these records are trusted in the hands of a single provider such as a healthcare centre or provider. Unfortunately, some of these healthcare centres suffer from various forms of attackers. According to the April 2019 Healthcare Data Breach Report by HIPAA [3], April 2019 has been one of the worse months for data breaches of healthcare records. Improper Disposal, Loss, Theft, Unauthorized Access/Disclosure and Hacking/IT Incident are reported to be the categories of causes of these breaches. according to [3] report. it goes without saying that traditional or existing data management systems need improvement or at least needs to some added layer security to help reduce or mitigate these data security breaches. Moreover, it is very difficult to assure transparency and data integrity with the existing traditional management system. For example, Patients record can be altered without the immediate awareness of the data owners. Imagine, a patient with HIV Negative status gets the record modified to HIV Positive by an unauthorized or malicious user. This can be detrimental and have dire consequence on the stakeholders of the records.

The aim of privacy-preservation is to provide protection of the privacy of the records during authorized disclosure and prevent unauthorized users from gaining access to the said records. The providers and data subjects, in this case, should have control over what, how and whom the records are shared with. This also forms the aim and goal of this paper.

In order to provide these security assurances, we designed a generic query authorization and authentication smart contract based on the Ethereum Blockchain technology.

The Blockchain technology, first proposed by [4] and used as a cryptocurrency, is a form of distributed ledger technology that stores transactions or records in blocks which are linked to each other in an incremental manner by the help of cryptography such that a change in a block invalidates the subsequent blocks. The network architecture of blockchain technology is decentralized and usually in a peer-to-peer form. Among the peers also known as nodes, there are special ones among them known as miners or block producers that are either nominated or qualify to write the transactions in the blocks and thereby get the other nodes to update their chain accordingly. How the selection is done is dependent on the underlying consensus mechanism adopted by the Blockchain in question. Bitcoin and Ethereum, for example, use Proof-of-Work (PoW) consensus mechanism. PoW requires the Miners to perform a specific task such as collecting and verifying the validity of the transactions on the network and attain a specified target hash value by hashing the collected transactions. The hash value of their blocks should be dependent on a previous block in order to make it valid. This effectively makes the Blockchain technology a temper-proof, difficult to take down by a DDOS attack and provide trustless and transparent integrity of the records stored on the network.

Ever since [4] use case was proposed by Satoshi Nakamoto, there have several variants and improvements of the technology of which a notable one Ethereum [5] Blockchain. The Ethereum Blockchain facilitates Smart Contract programming which is said to be a Turing complete.

Ethereum Smart Contracts enable programmable interaction with Blockchain technology. The Smart contract can sit between users and a traditional application and this is called a Decentralized Application (DAPP). What is special about Blockchain smart contract is that it is a trustless intermediary or if you like an API which is immutable once deployed and written to the blockchain. Ethereum provides the ability to run their technology either privately or rather make use of their Public Blockchain network. Our proposed model makes use of the Public Blockchain but can also be deployed on a private network if so desired.

**A. Technical challenges**

Blockchain technology has potentials to provide strong and relevant security to traditional systems but it has its own challenges. Generally, the technology suffers from majority or collision attacks. Well-known majority attack for the public Blockchain that utilizes Proof-of-Work (PoW) consensus mechanism is the 51% per cent attack. This refers to a miner or node owning more than 50% of the computing power of the network which gives it the capability of outperforming the other nodes in the mining process. The other majority attack is related to networks that depend on Proof-of-Stake (PoS). In that case, a miner or block producer who owns more stake or wealth on the network gets to monopolize the block production process. In any case, the nodes with such a power can behave arbitrarily to the detriment of the users of the Blockchain. Currently, the solution being relied upon is the huge cost involved in attempting such an attack. In other networks, the other users can also agree to penalize the offending node or block producer or simply decide to hard fork (branch) to a different chain on their own. salacity in terms of speed of the blockchain is also a challenge and till now does not really compete very well with traditional client-server systems. Researchers are working hard to address some of these challenges, and we intend contributing to this area in our future works as well.

The main contributions in this paper are: 1) Propose a pluggable Blockchain-based query authentication and authorization model, SmartCoAuth, to complement traditional systems for privacy-preservation and protection of data breach of sensitive records. 2) SmartCoAuth introduces query request token on the blockchain through a smart contract that occurs outside the data storage or record management system and thereby more secured and safer from common attacks on client-server architecture systems. 3) we focus our model on privacy-preservation of sensitive records when queried by external authorized users. In this case, their query tokens are created on the smart contract outside the record management system and then presented for verification before releasing such sensitive records to them. In addition, and unlike most other systems, our model also has the capability of providing control to selected stakeholders and on what to release, encrypt or not release at all.

## II. RELATED WORKS

Other researchers have also considered other techniques and security mechanisms to help address the privacy challenges faced by health records. Leveraging on the Mask Authenticated Messaging Module of IOTA DHL, [6] technology demonstrated its use with a wearable gadget such as Apple Watch that records personal health information in storing, sharing and retrieving such records. Their file format was based on the Fast Healthcare Interoperability Resources (FHIR) standards proposed by the Health Level Seven International (HL7) health-care standards organization. This is to enable the data to be sharable with existing systems that follow the standard. Their work effectively provides patient agency over the data to be shared. The patients are in full control of their records and can continuously share the data with a medical provider and revoke such sharing access at will.

Taking advantage of the Blockchain technology [7] proposed a blockchain-based framework called Ancile, by utilizing Ethereum smart contracts. Their framework provides secure, interoperable, and efficient access to medical records by patients, providers and third parties while preserving the privacy of the patient's sensitive information. Their goal was to analyse how their framework would be able to interact with the different needs of stakeholders and to understand how it could address the privacy and security concerns of the healthcare industry.

In order to meet the health records sharing requirements of the Office of the National Coordinator for Health Information Technology (ONC), [8] presented FHIRChain, a model based on Blockchain which uses the HL7 Fast Healthcare Interoperability Resources (FHIR) standard for sharing health records. They also developed a decentralized App (DAPP) using JavaScript and the Ethereum solidity smart contracts. The app was tested on the Ethereum testnet. Their proposed solution achieved several benefits such as increased modularity, fine-grained access control, identifiability & authentication and permission authorization. Effectively they were able to demonstrate data sharing by data owners without the need to upload or download data such as is the case of traditional server-client architecture.

On Personal health records (PHR) privacy preservation and sharing, [9] proposed OmniPHR, a blockchain-based model that integrates PHR for the use of both healthcare providers and patients. It effectively enables patients to have and maintain their health records in a unified environment from any gadgets of their choice and anywhere they find themselves. Healthcare providers benefit from the proposed solution by having their patient's records sharable among other providers.

## III. PRELIMINARIES

### A. Block production

Following the protocol of the network by performing special tasks and coming to an agreement on the final state of the blockchain by the participating nodes is known as Consensus Mechanism in Blockchain terminology. It is a core and most important aspect of Blockchain Technology. To this end, nodes in a PoW network compete to create blocks for other nodes to accept by verifying and approving transactions sent by users on the network after which they hash the results by including the hash of the previous block and a special integer known as the nonce. The final hash is supposed to be a value that is below or equal a target hash else the miner needs to keep trying different combination of the nonce and selected transactions until this target hash is attained. Finally, the miner sends this block to the other nodes or peers on the network for acceptance. It is only then the majority of the nodes accept this block and continue mining on top of it. Bitcoin and Ethereum use this kind of mechanism to reach consensus on the blocks. In other networks such as PoS networks, the block producers, in this case, do not have to perform mining activity i.e. compete in finding a target hash but rather stake their wealth and gain the right to create the next block on top of the current block. The requirement in all cases is to have the blocks link to previous ones. This link goes back to the first-ever block known as Genesis Block. The Genesis Block is fixed and given at the start or initial stage of the network in question.

Our model depends on the Ethereum Blockchain network hence rely on the consensus mechanism it utilizes. This means the success or failure of the Ethereum Blockchain affects our model accordingly.

B. **Ethereum Smart Contract**

Ethereum Blockchain allows a Turing complete kind of programs known as Smart Contracts to be deployed on the Blockchain and interacted with by external users. These smart contracts are executed as a special kind of transaction and are then mined or written to a block in the blockchain just like regular cryptocurrency transactions. After the deployment, external applications can call or execute functions in the smart contract which will intend present exactly what it is programmed to do. The Smart contracts operate in what is known as the Ethereum Virtual Machine (EVM). The smart contracts are usually written in a high-level programming language such as Solidity or Vyper and are then converted to Bytecode to be run on the EVM. Ethereum operates transactions and smart contract executions for that matter operates on the concept of gas. This implies that any function call, especially one that changes the state of the smart contract, consumes some amount of gas hence the need to pay for it and in the case of Ethereum, this payment is done with its cryptocurrency known as Ether. Our model requires a smart contract to be deployed in order to take advantage of the Blockchain technology.

## C. Signing transactions and messages

Blockchain technology relies heavily on public-key encryption for secure operation. Every user on the network, therefore, has public and private key pairs. The private keys are used to sign transactions to be sent to other users or smart contracts and the public keys are essentially the address to which they receive transactions from other users. Our model takes advantage of this to sign special secret messages needed for authentication and authorization.

## IV. PROPOSED MODEL: SMARTCOAUTH

### A. Overview

Our proposed model, SmartCoAuth, make use of one smart contract and a pluggable middleware module that sits between users' query requests and the database as well as capable of interacting with the deployed smart contract. The model also requires users to sign a secret message which they will have to present to the smart contract else their query will not have the necessary requirement for authorization. This secret message signing is done outside the Blockchain and the record management system by the user. The requirement for this is that the secret message must be signed with the corresponding private key of the public key or Ethereum address recognized by the record management system. It is worth mentioning that our architecture is flexible and can have several capabilities and features depending on the polices and security requirements of the provider. For example, the institution or provider can decide to submit encrypted records in a set to the query users until they get approve decryption key from the patients or data subjects. However, we keep our model simple and do not go to this length.

Fig. 1 outlines the various components that make all the SmartCoAuth architecture. It includes the Blockchain environment and the traditional records management system as well as the Smart Contract and the pluggable module.

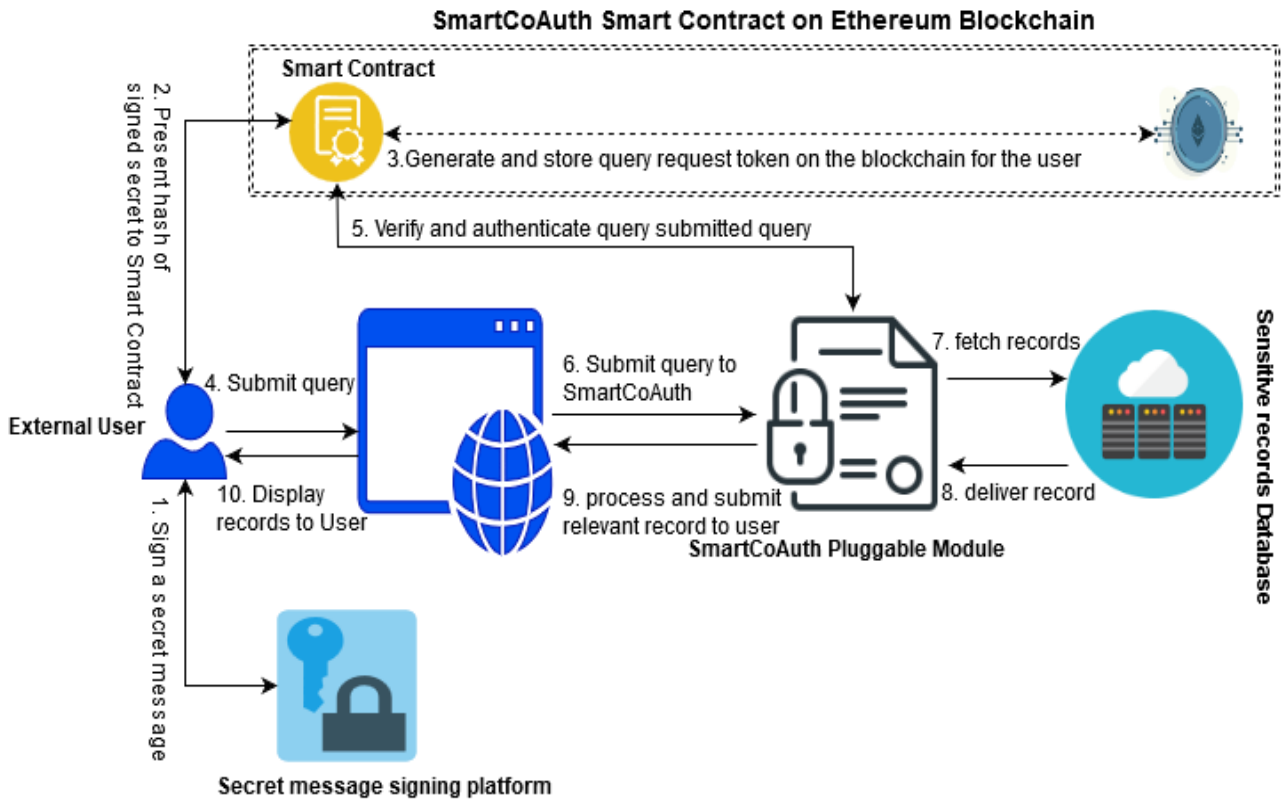

*Fig. 1 The schematic structure of SmartCoAuth Architecture with the various Components*

SmartCoAuth does not need the sensitive records to be stored in the Blockchain or smart contract as it is only serving as an authorization and authentication middleware for privacy-preservation. All that needs to be presented and stored in the smart contract is the hash of the signed secret message which is signed with the private key of the corresponding registered address. SmartCoAuth is then smart enough to compare the submitted secret along with the query and the hash stored in the smart contract in order to allow or deny access to the sensitive records. The query request and the secret message makes up the query token which is required by the pluggable middleware SmartCoAuth.

The main idea behind our model is providing a transparent, trustless and decentralized authentication and authorization mechanism which can also be audited by stakeholders to ascertain any possible abuse or even prevent one from happening. Due to the use of the Blockchain smart contract, there will be an absolute assurance that those who deserve to see the records are the ones who see it provided their private keys have not been stolen. And even in such a case, the use of their private keys by an attacker can easily be seen since the Smart Contract emits event logs that are permanently stored and can also be monitored remotely by the owner whenever a secret message is submitted on behalf of a registered query user.

We now proceed to outline the individual components of the proposed model and the function as well as workings of each of them. The overall algorithm pertaining to the middleware along with the rulesets will be discussed as well.

## V. Software components

SmartCoAuth is essentially made up of an Ethereum Smart Contract and A middleware that sits between a database and external user query requests. The model also requires query users to have access to a third-party and independent encryption system outside the blockchain and the records management system which they will need to create the hash of the encrypted secret message for submission to the smart contract.

### A. Smart Contract:

This is the Ethereum program deployed on the blockchain of which the users and the SmartCoAuth Middleware interact with. This is deployed on once and serves as the transparent and trustless platform for the stakeholders. In other words, all the users are aware of the logic and operation of the program and are confident about outcomes provided the end application utilizes it to provide the needed security.

It is made up of two main functions which we explain below:

*1) CreateQueryToken:*

This function registers the query request which will be required by the middleware to process the query. This function can be made public and open for users to call it without any notable security lapses. This is because SmartCoAuth middleware is smart enough to know whether the token was created by a registered user or not due to the signing of the secret message by the registered user. However, we make it public and have the owner account of the contract call it. This means users must interact with it using a dedicated page in our design

It takes in two parameters; the user address and the hash digest of the signed secret message. The smart contract then registers this and a block timestamp using mapping and array data types. Finally, it emits events which are logs that can be monitored by external applications and users whenever the function is called. This can help in alerting users, tracking and auditing of the query request tokens by stakeholders. In our design, we log the user address, the hash of the secret message and the block timestamp.

*2) getUserQueryToken:*

This function returns the query token whenever the SmartCoAuth middleware requires for it. It takes in the query user's address as its parameter and it is expected to return the address, the hashed message and the timestamp are found in the smart contract. The SmartCoAuth middleware uses this information to take further action depending on the outcome.

**B.** *SmartCoAuth Middleware:*

This is where the ruleset, algorithm and the decision to allow or deny a query request takes place. This sits between a database and query requests made by users.

Below are the rulesets for the possible query requests scenarios followed by the overall flowchart of the model designed to preserve the privacy of the sensitive data from external users.

*1) Overall flowchart*

Fig. 2 illustrates the overall process in our design by way of a flowchart:

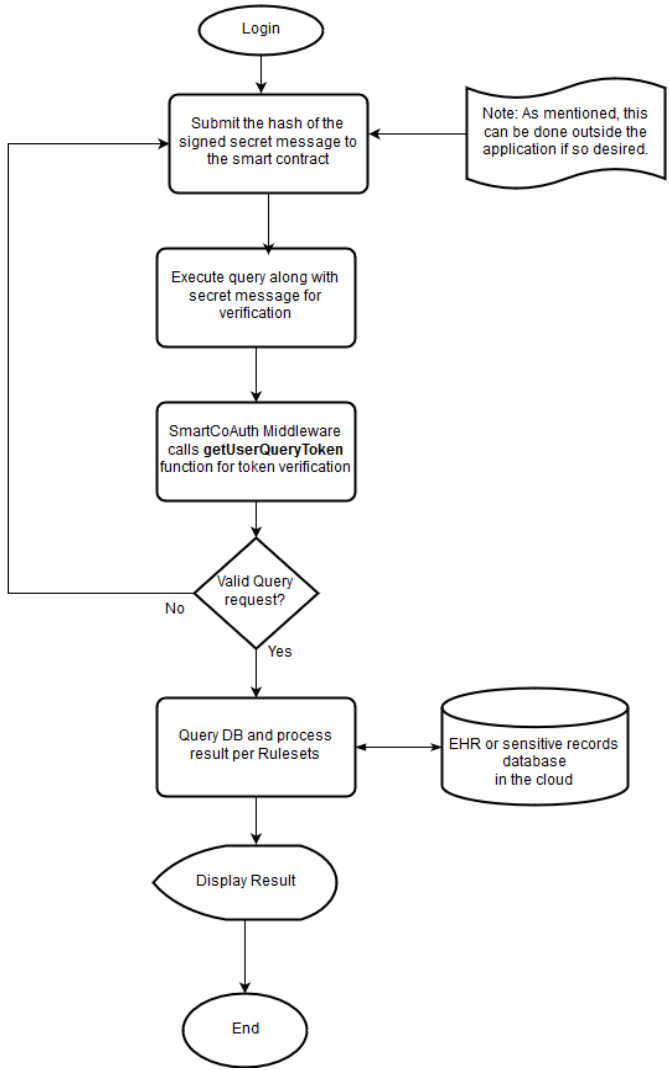

*Fig. 2 Flowchart of the proposed model*

*2) Ruleset*
The following are the proposed and possible rulesets (RS) that needs to be followed in our design:

   *a) RS1:*
User.Session = {Logged_In} && Query_Request_Token={Valid} && User.Status = {Authorized} -> Permission = {Allowed} U View.SensitiveRecords = {Full_Veiw_Access}

   *b) RS2:*
   User.Session = {Logged_In} && Query_Request_Token={valid} && User.Status = {Unauthorized} -> Permission = {Allowed} U View.SensitiveRecords = {Restricted_View_Access}

   *c) RS3:*
   User.Session = {Logged_In} && Query_Request_Token={Invalid} && User.Status = {Authorized} -> Permission = {Allowed} U View.SensitiveRecords = {Restricted_View_Access}

   *d) RS4:*
   User.Session = {Logged_In} && Query_Request_Token={Invalid} && User.Status = {Unauthorized} -> Permission = {Allowed} U View.SensitiveRecords = {Restricted_View_Access}

   *e) RS5:*

User.Session = {Not_Logged_In} -> Permission = {Anonymous} U View.SensitiveRecords = {Anonymized_View || Access_denied}

As can be observed from the above rulesets, the only user who in our design qualifies to have full access to the query result is the authorized logged in user with a valid query request token. All other condition gets restricted view access based on the policy of the organization in question. Example of such restricted view access could be to obfuscate some key fields such as the ID, NAME, DATE OF BIRTH, MEDICAL CONDITION etc and/or only give aggregate result instead.

C. *Database server:*

This is where the sensitive records are stored and it can be any traditional database management system as PostgreSQL, SQL Server, MYSQL, MongoDB etc. The requirement in our model is that every recordset from here to the frontend users will have to pass through SmartCoAuth middleware. This server also stores the public key or blockchain addresses of the registered and authorized users even though our architecture does not necessarily require this to be the case. The addresses can be stored in another location if preferred provided it is reachable by SmartCoAuth middleware. It is also worth noting that, due to the flexibility of our model encrypted version of sensitive fields or records can be stored here instead and get decrypted on the fly if so desire. In the same vein, hashes of the addresses can also be stored for confidentiality purposes and if this is done then users provide their registered addresses for SmartCoAuth to verify before granting them the needed access. We, however, keep our model simple in this paper without hashing or encrypting the records stored in the database.

D. *External Encryption Manager:*

This is the external secret message signing for the users and it is independent of the blockchain and the records management system. In our case, we use the Ethereum remix platform (https://remix.ethereum.org) for this purpose but any other suitable application can be used.

E. *Ethereum Blockchain:*

Our design depends on the Ethereum Public Blockchain and all its security and performance provision. However, it is possible to have a private Blockchain set up using the Go-Ethereum (Geth). In that case, the security and maintenance of the Blockchain and the consensus mechanism must be decided and taken care of by the institution.

## VI. EXPERIMENT AND RESULTS ANALYSIS

The proposed solution was simulated on a Windows 10 desktop Computer using Python and Web3.py API. Python's Django Web application framework was used to develop the Healthcare system with all the needed features and proposed ruleset and algorithm. The Web3.py library was used to implement the SmartCoAuth Middleware features that provide the added layer security between the database and the queries from the external users at the server-side. Web3.js was used to provide the client-side smart-contract interaction. It is important to mention that the server-side interaction of the smart contract with the Web3.py is not essential in our proposed model if users are made to interact with the smart contract on their own instead of doing so from the application. The web3.js JavaScript version can be used with any programming to implement the functionality of the middleware.

We simulated the Blockchain network with Ganache, an Ethereum blockchain client for development. The dataset contained 12500 records with 8 attributes which when put together form the sensitive record of the data subjects. We measured the average processing time of the queries per record and compared it to scenarios without our proposed added layer security.

### A. *Threat Model and Privacy-Preservation security analysis*

#### 1) *Stolen Private Key*

If a query request user loses his private-key and it used to create a query request token on his behalf this can result in impersonation attack and consequently result in data being leaked to an unauthorized user. In this case, the security provided by our model will be the events that are fired whenever users submit the secret message to the smart contract. Users are expected to monitor this event and report accordingly in order to prevent this attack. With SmartCoAuth approach, users can be asked to provide a special kind of secret message pattern which is only known to them. If this is the case, then it means the attacker also needs to know this patter besides knowing the private key of the user since the token will be invalid otherwise.

In addition, since our model is pluggable to traditional application and acts as an added layer of security this also means that the attacker can only succeed if he has the username and password of the authorized user.

*2) Replay attack*

This is when the attacker attempts to use another user's already signed secret message for his own session. This will fail with SmartCoAuth because the logged-in user's Ethereum Address is known by SmartCoAuth Middleware and will attempt to verify the signed message with that address. If there corresponding private key was not used to sign this secret message, then it will fail. The only way the attacker will succeed in this case is to also manage to change his registered Blockchain address to match the right user. Changing address in this manner is assumed under our model to be impossible or not easy to do especially also because that field is a unique field and there should not be more than one user sharing the same blockchain address. In addition, the database administrators also need to approve registered address or changes made by users.

We also recommend regular expiration of the query token for better security.

*3) Guess secret message*

If an attacker manages to guess the secret data, then he also needs to log in as the user in question in order to succeed with the attack. This secret data signed by the user should not be known by anyone in the first place. A regular expiration of the token and avoidance of secret data re-use can help prevent this attack.

**B.** *Performance Analysis*

We compared SmartCoAuth average query processing time as the number of results set increases with an approach such as [10] that depends on only the ruleset without contacting the blockchain or a smart contract.

As can be observed in Figure 3, our approach entails a delay of about 25ms on average to return the results set than a regular approach such as [10] which does not depend on the Blockchain. This makes sense because SmartCoAuth middleware first needs to contact the smart contract before fulfilling any query request for improved security. This delay is, therefore, dependant on the performance of the Blockchain and how fast it returns results when smart contract functions are called. Even though there is this amount of delay it is worthwhile trade-off due to the added security that ensues from using it.

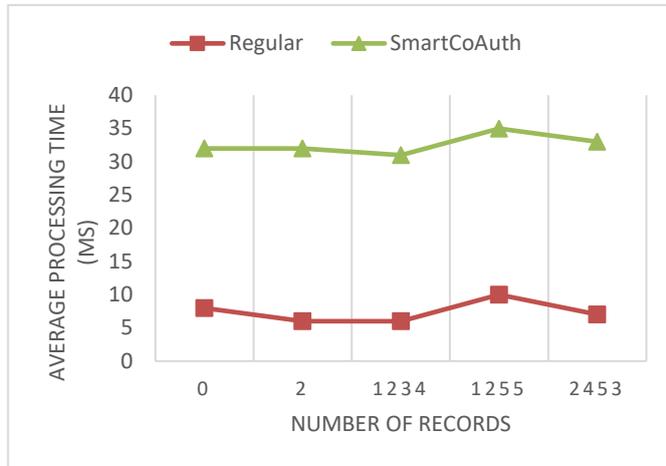

*Figure 3 Average query processing time with respect to returned results*

### C. *Privacy preservation*

SmartCoAuth by design has several dimensions for privacy-preservation. By registering every query request token on the Blockchain and subsequently verifying it on the blockchain before approving them, stakeholders can rest assured that whoever has queried their database for sensitive records have the required and legitimate right to do so. More so, such request and fulfilment are monitored and tracked in a trustless environment such as the blockchain. Privacy also involves control of users as to who, when and how their information is shared [1]. SmartCoAuth can also facilitate this by having by providing encrypted results and then after have the data owners or authorized administrators provide special one-time decryption keys to the users who make the request. All these procedures can be executed and monitored on the blockchain. In the case of data being leaked, it will be easy to identify the offending party. The use of public-key cryptography in SmartCoAuth ensures non-repudiation security objective.

However, even though SmartCoAuth can provide an added layer of security to existing traditional systems, any major flaw of that traditional system can affect its security provision. For example, if an attacker can compromise the database records at the database layer then SmartCoAuth middleware cannot identify this form of attack and will therefore not be able to track any data leak as a result. This means the security assurance of SmartCoAuth depends heavily on having the query request pass through the middleware for verifying, authentication and authorization.

The use of Blockchain technology cannot be understated because records kept there are not changeable and cannot be tampered with once confirmed in a block and accepted by the network. This can be of tremendous help during auditing. However, it is also important to note that storing too much information on the public and permissionless blockchain can also have some adverse effect because it can also lead to statistical analysis attacks and information gathering that can result in disclosure of valuable information to an attacker. Having users remaining anonymous or pseudonymous behind their public keys or Ethereum address can help to mitigate this.

## VII. CONCLUSION

In this paper, we designed an added layer security model, SmartCoAuth, based on the Blockchain technology that can be implemented on to of traditional records management systems. Our model enables a transparent authorization, authentication and accountability of sensitive query requests by external users. The model can securely and in a trustless manner fulfil query requests such that the actions are monitored and audited by all stakeholders involved and thereby provide a transparent and tamper-resistant of who, how, when and what has been queried by external users. SmartCoAuth use of Smart Contract provides high-level trust of the privacy-preservation mechanism.

SmartCoAuth was designed to complement existing traditional client-server architecture solutions. The use of encryption and hashing of data on the blockchain and/or the delivered records gives the assurance of non-repudiation and effective auditing and monitoring on the Blockchain.

It is, however, important to note that the use of the Smart Contract implies that the success and efficiency of SmartCoAuth also depend on the success of the underlying Blockchain technology. We observed the overhead in query execution when using SmartCoAuth but we believe it is worth it given the nature of security provided by the Blockchain technology. We do not, therefore, suggest SmartCoAuth be the silver bullet to the impending and worrying data breaches over sensitive records such as healthcare records stored in the cloud. However, we continue to explore and research in this area for better ways and more efficient ways to meet data privacy security requirements in the industry.